\documentclass[12pt]{iopart}
\usepackage{graphicx}
\usepackage[english]{babel}
\usepackage{amssymb}
\usepackage{multirow}
\usepackage{array}
\usepackage{url}

\begin{document}

\title[Sensitivity of YAC  to observe the light-component spectrum of
primary cosmic rays]
{Sensitivity  of YAC  to measure the light-component spectrum of
primary cosmic rays at the ``knee" energies}

\author{L~M~Zhai$^1$, J~Huang$^1$, D~Chen$^2$, M~Shibata$^3$, Y~Katayose$^3$, Ying~Zhang$^1$, J~S~Liu$^1$, Xu~Chen$^1$, X~B~Hu$^{1,4}$ and Y~H~Lin$^1$}
\vspace{10pt}

\address{$^1$ Key Laboratory of Particle Astrophysics, Institute of High Energy Physics, Chinese Academy of Sciences, Beijing 100049, China}
\address{$^2$ National Astronomical Observatories, Chinese Academy of Sciences, Beijing 100012, China}
\address{$^3$ Faculty of Engineering, Yokohama National University, Yokohama 240-8501, Japan}
\address{$^4$ Department of Physics, Shandong University, Jinan 250100, China}

\ead{huangjing@ihep.ac.cn}
\vspace{10pt}
%\begin{indented}
%\item[]February 2014
%\end{indented}

\begin{abstract}

 A new air-shower core-detector array (YAC : Yangbajing Air-shower Core-detector array)  has been developed to measure the primary cosmic-ray composition at the ``knee" energies in Tibet, China, focusing mainly on the light components. The prototype experiment (YAC-I) consisting of 16 detectors has been constructed and operated at Yangbajing (4300 m a.s.l.) in Tibet since May 2009. YAC-I is installed in the Tibet-III AS array and operates together. In this paper, we performed a Monte Carlo simulation to check the
sensitivity of YAC-I+Tibet-III array to the cosmic-ray light component of cosmic rays  around the knee energies, taking account of the observation conditions of actual YAC-I+Tibet-III array. The selection of light component from others was made by use of an artificial neural network (ANN).
The simulation shows that the light-component spectrum estimated by our methods can well reproduce the input ones within 10\% error, and there will be about 30\% systematic errors mostly induced by the primary and interaction models used.
It is found that the full-scale YAC and the Tibet-III array is powerful to study the cosmic-ray composition, in particular, to obtain the energy spectra of protons and helium nuclei around the knee energies.

\end{abstract}

\pacs{98.70.Sa, 96.50.sb, 96.50.sd}
\vspace{2pc}
\noindent{\it Keywords}: cosmic rays, hadronic interaction, knee, composition, energy spectrum

\submitto{\jpg}
\maketitle

\section{Introduction}
\label{intro}

It is well known that the all-particle spectrum of primary cosmic rays follows a power law of  dJ/dE $\propto$ E$^{-\gamma}$, but steepens at energies around 4$\times$10$^{15}$ eV where the power index $\gamma$ changes sharply from $\sim$2.7 to $\sim$3.1~\cite{Amenomori-2008,Horandel-2003}.
Such structure of the all-particle energy spectrum is called the ``knee", which is considered to be closely related to the origin, acceleration and propagation mechanism of cosmic rays. In order to explain the existence of the knee, many hypotheses and mechanisms~\cite{Horandel-2004,Shibata-2010} have been proposed. Although all these approaches can well describe the knee structure, there are much discrepancies in the prediction of the individual components at the knee region. Thus, it is critical to measure the primary chemical composition or mass group at energies 50-10\,000 TeV, especially, to measure the primary energy spectra of individual component and determine a break energy of the spectral index for individual  component.
Direct cosmic-ray measurements on board balloons or satellites are the best way to study the chemical composition, while the maximum energy they can cover is up to 10$^{14}$ eV/nucleon at most  due to limited detection area or exposure time. We may have no choice but to rely on ground-based air-shower (AS) measurements to study the primary chemical composition around the knee.

The study of cosmic-ray composition around the knee was done by a hybrid experiment of the emulsion chambers (ECs), the burst detectors (BDs) and the AS array (Tibet-II), where ECs and BDs of total area 80 m$^2$ were set up near the center of the AS array and operated for three years~\cite{Tibet-EC-2000a,Tibet-EC-2000b,Tibet-EC-2006}. The threshold energy of ECs capable of analyzing the fine structure of AS cores is about 1 TeV, so that it is not difficult to separate the AS events induced by light-component of protons and helium nuclei, while the energy range of primary particles is limited to be above $\sim$200 TeV for protons and $\sim$400 TeV for helium nuclei~\cite{Tibet-EC-2000b}. This experiment suggests that the flux of light component is less than $\sim$30\% of the total, resulting in that the knee is dominated by nuclei heavier
than helium~\cite{Tibet-EC-2006}. A demerit of this experiment is that there
are few statistics of the high-energy core events due to the high detection threshold energy of the experiment as mentioned above. To improve this problem,  a new air-shower core detector named YAC (Yangbajing Air shower Core detector) has been developed and improved so as to meet our requirements.

One important improvement is to lower the detection threshold energy of primary particles to several times 10 TeV, about one order of magnitude smaller than the previous experiment. With this improvement, the energy spectra of individual components measured by YAC will overlap with those of direct measurements, which may help us to examine  the knee of light component, such as ``proton knee" or ``helium knee".
Another important improvement of YAC is its ability to count the number of shower particles passing through each detector in a wide dynamic ranging  from 1 to 10$^6$ particles, making it possible to observe the primary cosmic rays
in the energy range from $\sim$10 TeV to $\sim$10 PeV.

Until now, we have constructed and operated  YAC-I as a prototype of full-scale
YAC comprising 400 core detectors. YAC-I  is a small array consisting of 16 core detectors which were placed near the center of the Tibet-III AS array as shown in Fig.~\ref{figure-01}, while being able to observe a lot of AS-core events in the energy around $10^{14}$ eV by the operation of a few months. As the primary
composition around this energy region is fairly well known by the direct
observations~\cite{ATIC2-2005,CREAM-nuclei,CREAM-PHe}, the data from YAC-I may be used to test the interaction models such as SIBYLL2.1, QGSJETII-04 and EPOS-LHC being currently used in the Monte Carlo (MC) simulations.  In this paper, we discuss the performance and
sensitivity of YAC for observing light-component spectrum of primary particles
through MC simulations based on the YAC-I experiment.

\section{YAC-I Experiment}
\label{expt}

\begin{figure}[!ht]
\centering
\includegraphics[width=0.8\textwidth]{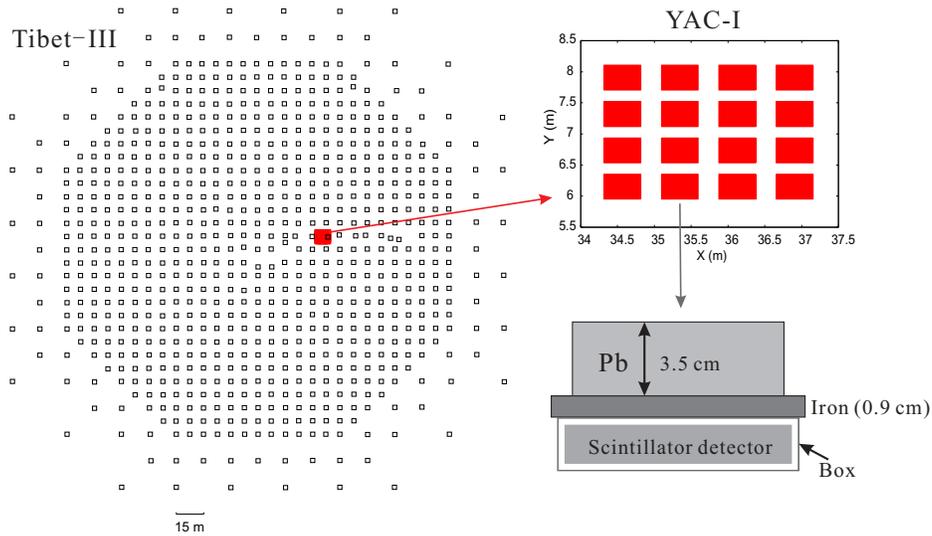}
\caption{Schematic view of YAC-I and Tibet-III AS array. Open squares : scintillation detectors of the Tibet-III array. For details of the Tibet-III, see the paper~\cite{Amenomori-2008}. Filled red squares : core detectors of the YAC-I array. YAC-I consists of 16 detector units of 0.2 m$^2$ each, which cover the area of about 10 m$^2$. Each detector of YAC-I consists of a lead plate with the thickness of 3.5 cm (6.3 radiation lengths), a supporting iron plate with the thickness of 0.9 cm (0.5 radiation lengths) and a plastic scintillator with the thickness of 1 cm.}
\label{figure-01}
 \end{figure}

\begin{figure}[!ht]
  \centering
  \includegraphics[width=0.7\linewidth]{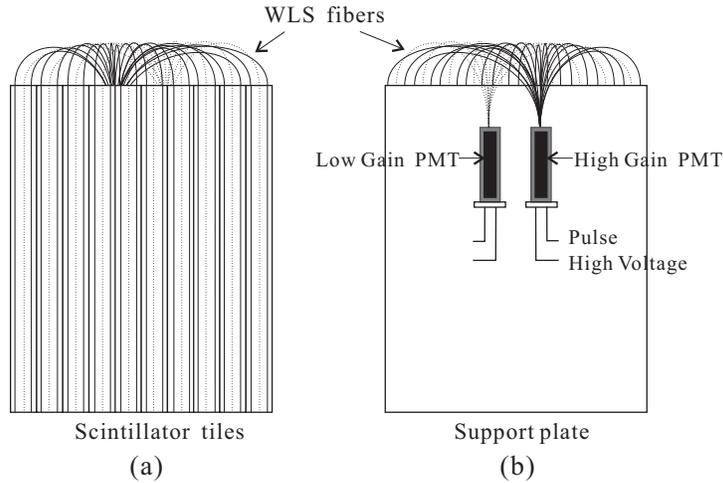}
  \caption{ Schematic view of the scintillation detector, the top view (a) and the back view (b). The scintillator is divided into 10 pieces of 4 cm width and scintillation lights from each piece are collected through the wavelength shifting (WLS) fibers. }
  \label{figure-02}
 \end{figure}

The YAC-I array consists of 16 core detectors which are placed on a
4$\times4$ square grid covering the area of about 10 m$^2$ as shown in Fig.~\ref{figure-01} and has been operating since May 2009, together with the Tibet-III AS array. Each core detector of YAC-I comprises a plastic scintillator with the size of
40 cm $\times$ 50 cm and a lead plate with the thickness of 3.5 cm (6.3 radiation lengths) being put on the scintillator. The lead plate is used to select AS particles and cores capable of having sufficient energy to create cascade showers in the lead plate and pass through the scintillator.
The plastic scintillator in the core detector is divided into 10 pieces of the width of 4 cm and the scintillation lights are collected through wavelength shifting (WLS) fibers as shown in Fig.~\ref{figure-02}. Such design ensures the geometrical uniformity of detector response within 5 \%. The details about the hardware of YAC detector is described in~\cite{YAC-detector-jiang, Hardware-YAC-Liwj}. Two photomultiplier tubes (PMTs) of high-gain (HAMAMATSU: R4125) and low-gain (HAMAMATSU: R5325) are equipped  to cover  a wide dynamic range from 1 MIP(Minimum Ionization Particle) to 10$^6$ MIPs  as seen in Fig.~\ref{figure-02}. The corresponding linearity and saturation effect of PMT and scintillator were examined by use of  cosmic-ray muons and electron beams provided by the beam facility of BEPCII (Beijing Electron Positron Collider, IHEP, China)~\cite{Hardware-YAC-chen}. The stability of the PMT gain was  checked and corrected using cosmic-ray muons.

 In this experiment, the YAC-I array works to observe air-shower cores and their  accompanying ASs are observed simultaneously with the Tibet-III AS array. The Tibet-III provides information on the arrival time and direction of each air shower and its AS size corresponding to the energy of primary particle~\cite{Amenomori-2008}.
 When a YAC-I event is triggered, its accompanying AS is simultaneously recorded by the Tibet-III array and  the matching between YAC-I and Tibet-III events is made by their arrival time stamps recorded by a GPS clock.

\section{Monte Carlo Simulation}
\label{mc}

We have carried out a full  Monte Carlo (MC) simulation on the development of air showers in the atmosphere  using the simulation code Corsika~\cite{Heck-Corsika} (version 7.3500).
Three hadronic interaction models, including SIBYLL2.1~\cite{siby}, EPOS-LHC (v3400)~\cite{eposlhc} and QGSJETII-04~\cite{qgs}, are used to generate the air-shower events in the atmosphere.
For the primary cosmic rays, we examined three composition models, namely, ``He-poor", ``He-rich" and ``Gaisser-fit" models, in order to evaluate the systematic errors attributable to primary composition models.
The ``He-poor" model is based on the HD (Heavy Dominant) model mentioned in the paper~\cite{Amenomori-2008},
and it is slightly revised to match with the new all-particle energy spectrum~\cite{Amenomori-2008}, to be
the new ``He-poor" model. The ``He-rich" model is the ``Model B (a lightly harder spectrum than the previous on by taking account of the nonlinear effects)" mentioned in the paper~\cite{Shibata-2010}. The ``Gaisser-fit" model is the ``three-population" model mentioned in the paper~\cite{Gaisser-model}. The proton spectra of the former two models are fitted to the direct measurements at the low energy and consistent with the spectrum obtained from the Tibet-EC experiment at the high energy. The He spectrum of He-poor model coincides with the results from RUNJOB, but the He spectrum of He-rich model coincides with the results from JACEE, ATIC2 and CREAM. The Gaisser-fit model fits to a higher He model (almost same as our He-rich model) at the low energy range and to the KASCADE-QGSJET data at high energy range in which light components (P and He) dominate in the chemical composition. In all models mentioned above, each component is summed up so as to match with the all-particle spectrum with a sharp knee, which was obtained with the Tibet-III AS array~\cite{Amenomori-2008}.

\begin{figure}[]
  \centering
  \includegraphics[width=0.76\textwidth]{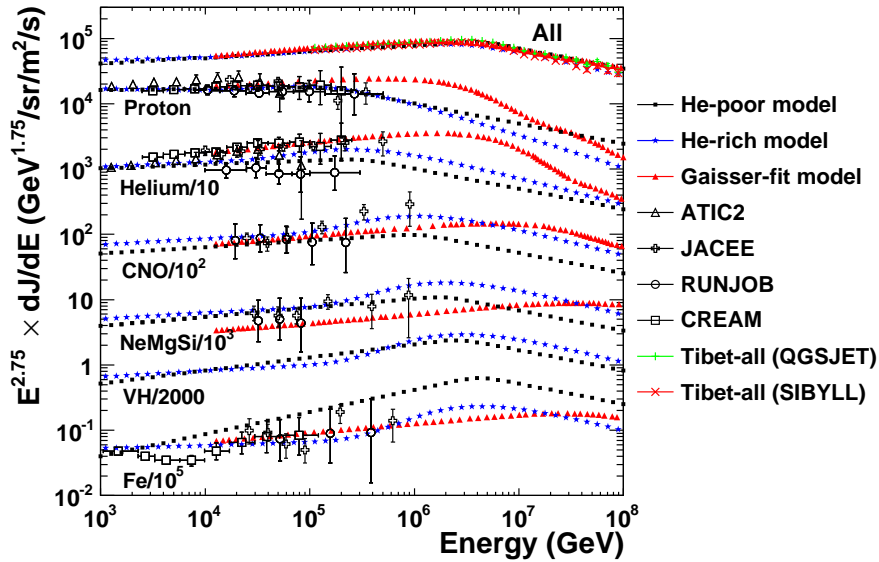}
  \caption{Primary cosmic-ray composition for He-poor, He-rich and Gaisser-fit models compared with those  of direct measurements (ATIC2~\cite{ATIC2-2005}, JACEE~\cite{JACEE-1998}, RUNJOB~\cite{RUNJOB-2001}, CREAM~\cite{CREAM-nuclei,CREAM-PHe}) and the sum of all components (all-particle spectrum) compared with the results obtained by the Tibet-III experiment~\cite{Amenomori-2008} .}
  \label{figure-03}
 \end{figure}

Table~\ref{table-1} is a summary of the fractions of the components (P, He, Medium and Fe) of the three composition models in given energy regions for three primary models. The energy spectra of individual components (or mass groups) for three primary models are shown in Fig.~\ref{figure-03}. It is seen that all the individual components of the three models in the low energy range (less than 100 TeV) are in good agreement with direct measurements while differ significantly at higher energy.
The all-particle spectra of three models, however, coincide with each other and reproduce the sharp knee structure as well.

\begin{table}[!t]
\caption{The fractions of individual components in the assumed primary cosmic-ray spectra of  He-poor, He-rich and Gaisser-fit models. }
\label{table-1}
\begin{indented}
\item[]
\begin{tabular}{@{}llccc}
\br
%-------------------------------------------------------------------------------------------------------------------------------
  Composition  &\multirow{2}{*}{Components}     & $10^{13}-10^{14}$ eV   & $10^{14}-10^{15}$ eV   & $10^{15}-10^{16}$ eV   \\
  Models       &                               &(\%)                    &(\%)                    &(\%)                    \\
\mr
%-------------------------------------------------------------------------------------------------------------------------------
\multirow{4}{*}{He-poor}     & P         & 31.5        & 22.7        &9.6         \\
                             & He         & 22.4        & 18.8        &9.7         \\
                             & Medium      & 26.6        & 26.6        &26.1        \\
                             & Fe           & 19.5        & 31.9        &54.6        \\
\mr
%-------------------------------------------------------------------------------------------
\multirow{4}{*}{He-rich}     & P        & 31.1        & 26.3        &10.0         \\
                             & He        & 25.1        & 28.7        &17.5         \\
                             & Medium     & 32.4        & 34.4        &50.3         \\
                             & Fe          & 11.3        & 10.6        &22.2         \\
\mr
%-------------------------------------------------------------------------------------------
\multirow{4}{*}{Gaisser-fit}   & P         & 32.8        & 29.0        &19.6        \\
                               & He         & 34.4        & 37.4        &37.4        \\
                               & Medium      & 20.1        & 20.4        &25.3        \\
                               & Fe           & 12.7        & 13.2        &17.7        \\
\br
%--------------------------------------------------------------------------------------------------------------------------------
\end{tabular}
\end{indented}
\end{table}

In this simulation, primary cosmic rays at the top of the atmosphere within the zenith angles smaller than 60 degrees are thrown  into the atmosphere isotropically and the minimum energy of primary cosmic rays is set to 40 TeV.
All shower particles in the atmosphere are traced down to the minimum energy of 1 MeV. The AS events generated  are randomly dropped onto the area  of 32.84 m $\times$ 32.14 m, which is a 15 m wider in each side of the YAC-I array.
This dropping area is large enough to collect  the AS events more than 99.5\%  under our core-event selection conditions
(see below in the text). Observation of the MC events is made with  the same method as that of the experiment.

The detector responses to shower particles falling on the detectors of (YAC-I+Tibet-III) array are calculated using the Geant4~\cite{Geant4} (version 9.5), where the detector performance, trigger efficiency and effective area are adequately taken into account based on the experimental conditions.
The number of charged particles passing through the scintillator is defined as the PMT output (charge) divided by that of the single-particle peak. The single-particle peak is determined by a probe calibration~\cite{Amenomori-2008, YAC-detector-jiang} using cosmic rays, typically  muons. The value of single-particle peak is measured as 1.98 MeV for YAC-I detectors and 6.28 MeV for the Tibet-III detectors. These values are used  in this MC simulation.

The main purpose of this work is to check the sensitivity of YAC to observe the light component of primary cosmic rays as well as to evaluate the systematic errors by adopting different primary composition models and interaction models mentioned above.
For this, we selected five combinations of interaction models and primary composition models. Four combinations of SIBYLL2.1+He-rich, SIBYLL2.1+He-poor, EPOS-LHC+He-poor and EPOS-LHC+Gaisser-fit are to check the sensitivity of YAC to the light component and uncertainties due to the adoption of the different composition models. Other three combinations of SIBYLL2.1+He-poor, QGSJETII-04+He-poor and EPOS-LHC+He-poor are to check the interaction models and also uncertainties under the same primary composition model. It has value to point out here that there is no serious difference among the current interaction models on the particle production in the forward region and proton-air inelastic cross sections in our concerned energy region from 10 TeV to $10^4$ TeV since all the models are well tuned using recent accelerator data including LHC, while there are big differences among primary composition models because of a lack of direct observation data at energies above $\sim$200 TeV.

The number of air-shower events generated for each  model is 7.40$\times$10$^{7}$, 6.57$\times$10$^{7}$, 4.67$\times$10$^{7}$, 6.25$\times$10$^{7}$ and 5.18$\times$10$^{7}$, respectively, as shown in Table~\ref{table-2}.
The analysis of these MC events was made by the same method used in the experiment.

\section{Analysis}
\label{analysis}

Information on the size $N_e$ and arrival direction of each air shower event hitting both YAC-I and Tibet-III arrays can be easily obtained from the MC events observed with the Tibet-III AS array simultaneously. Details of its analysis are found in the paper~\cite{Amenomori-2008}.

From the YAC-I array, we can obtain the following five quantities reflecting the characteristic of AS cores : (1) ${N_{hit}}$, the number of ``fired'' detectors with ${N_b}$$\geq$ 200, where $N_b$ is the number of particles (burst size) observed
by each core-detector  ; (2) $\sum$${N_b}$, the total sum of $N_b$ of fired detector ;
(3) ${N_b}$$^{top}$, the maximum ${N_b}$ among the fired detectors ;
(4) $\langle R \rangle$, the mean lateral spread defined as $\langle R \rangle$ = $\sum$${r_i}$/($N_{hit}$-1) ; (5) $\langle N_bR \rangle$, the mean energy-flow spread defined as $\langle N_bR \rangle = \sum ({N_b^i} \times {r_i})/N_{hit}$,
where $N_b^i$ denotes the number of particles observed in  $i$-th fired detector and $r_i$ represents the lateral distance from the burst center ($X_c, Y_c$), where ($X_c, Y_c$) = $\left(\frac{\sum N_b^ix_i}{\sum N_b^i}, \frac{\sum N_b^iy_i}{\sum N_b^i}\right)$.
It is confirmed that the five quantities mentioned above are basic and enough to separate the light component (P+He) from others. A use of ANN-method~\cite{ANN} may further improve the quality of separation.

\begin{table}[!ht]
\caption{Statistics of the data sets selected in MC simulation. }
\label{table-2}
\begin{indented}
\item[]
\begin{tabular}{@{}lcc}
\br
%-------------------------------------------------------------------------------------------------------------------------------
  \multirow{2}{*}{Models}          &Primaries                       &Core events                \\
                                   &(E $\geq$ 40 TeV)               &(Mode energy: $\sim$200 TeV)   \\
\mr
%-------------------------------------------------------------------------------------------------------------------------------
  SIBYLL2.1+He-rich                & 7.40$\times$10$^{7}$           & 64\,331          \\
  SIBYLL2.1+He-poor                & 6.57$\times$10$^{7}$           & 47\,580          \\
  QGSJETII-04+He-poor              & 4.67$\times$10$^{7}$           & 31\,928          \\
  EPOS-LHC+He-poor                 & 6.25$\times$10$^{7}$           & 42\,137          \\
  EPOS-LHC+Gaisser-fit             & 5.18$\times$10$^{7}$           & 49\,390          \\

\br
%-------------------------------------------------------------------------------------------------------------------------------
\end{tabular}
\end{indented}
\end{table}

\begin{figure}[!ht]
  \centering
  \includegraphics[width=0.56\linewidth]{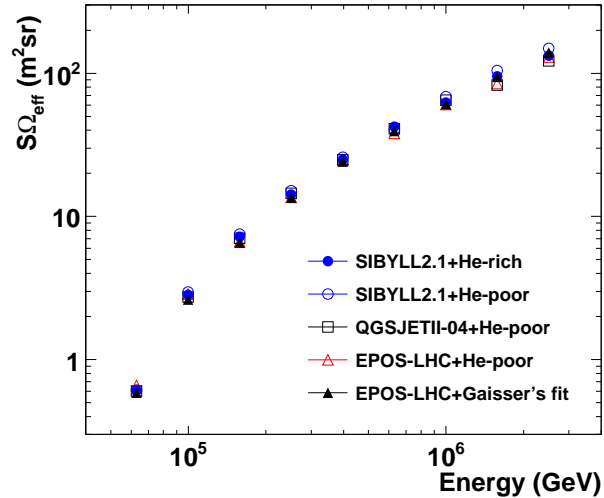}
  \caption{ The effective $S\Omega$ of YAC-I array to observe the light-component for various models used in MC.}
  \label{figure-04}
 \end{figure}

In order to obtain the light-component spectrum using the data from
both arrays of YAC-I and Tibet-III, we select the high-energy core events by
imposing the  conditions of  ${N_b}\geq 200$, $N_{hit}\geq4$, ${N_b}^{top}\geq1500$ and ${N_e}\geq80\,000$.
The mode energy of primary particles producing such
high energy core events is then estimated to be about 200 TeV.
The statistics of the data-sets selected based on the five models are listed in Table~\ref{table-2}. Shown in Fig.~\ref{figure-04} is
the effective $S\Omega$ of YAC-I array to observe the AS-core events satisfying the event select conditions, where $S$ denotes the detection area and  $\Omega$ the solid angle. The effective $S\Omega$  depends weakly on the model
used, but its difference is found to be smaller than 25\% in our concerned energy range.

\section{Results and Discussion}
\label{results}

We check the sensitivity of YAC array to the interaction models and primary
cosmic-ray models using  the high-energy core events selected under the
conditions discussed in the previous section.

\subsection{Total burst-size  spectrum and  mean lateral spread of AS-cores}

\begin{figure}[!ht]
  \centering
  \includegraphics[width=0.48\linewidth]{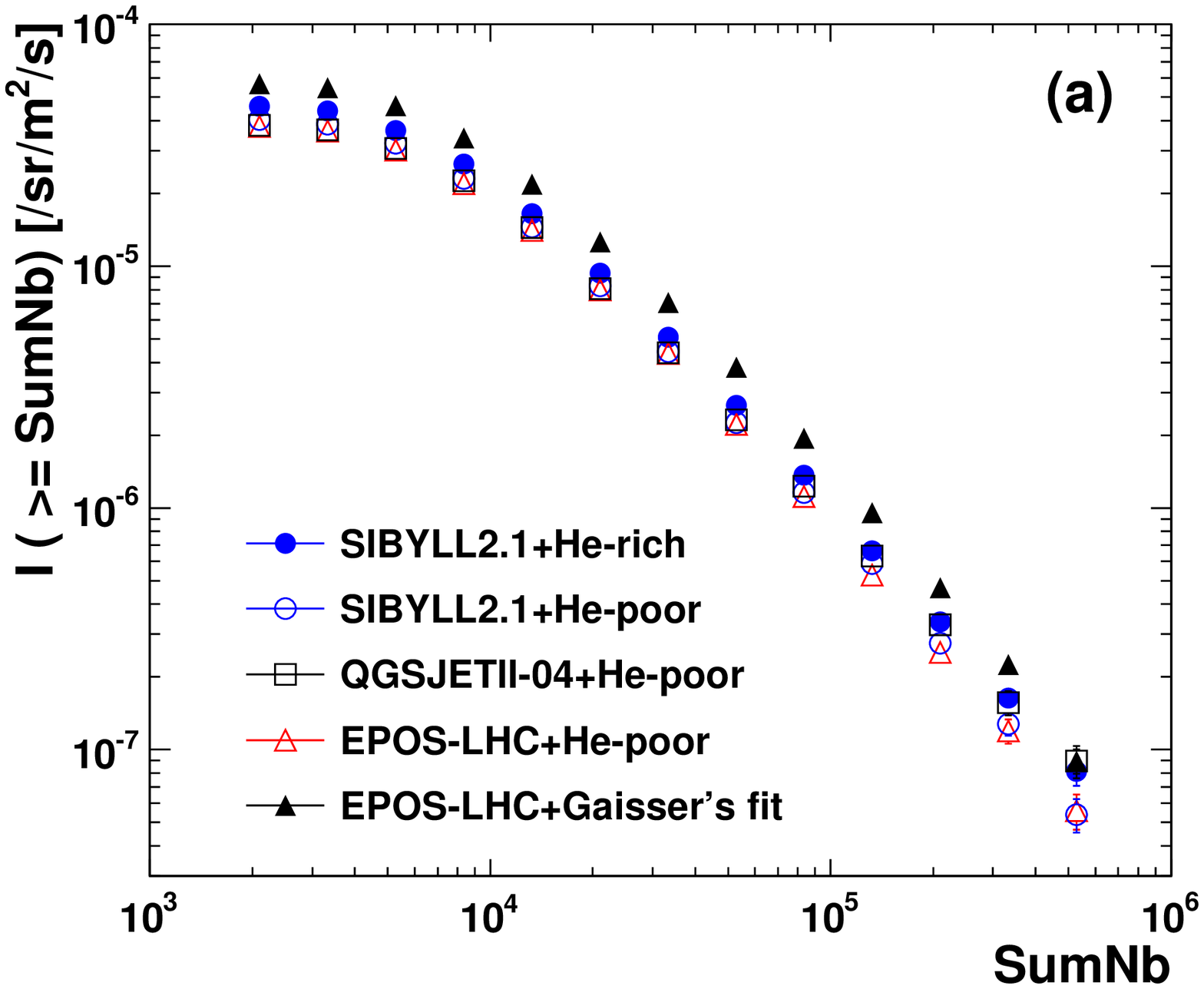}
  \includegraphics[width=0.48\linewidth]{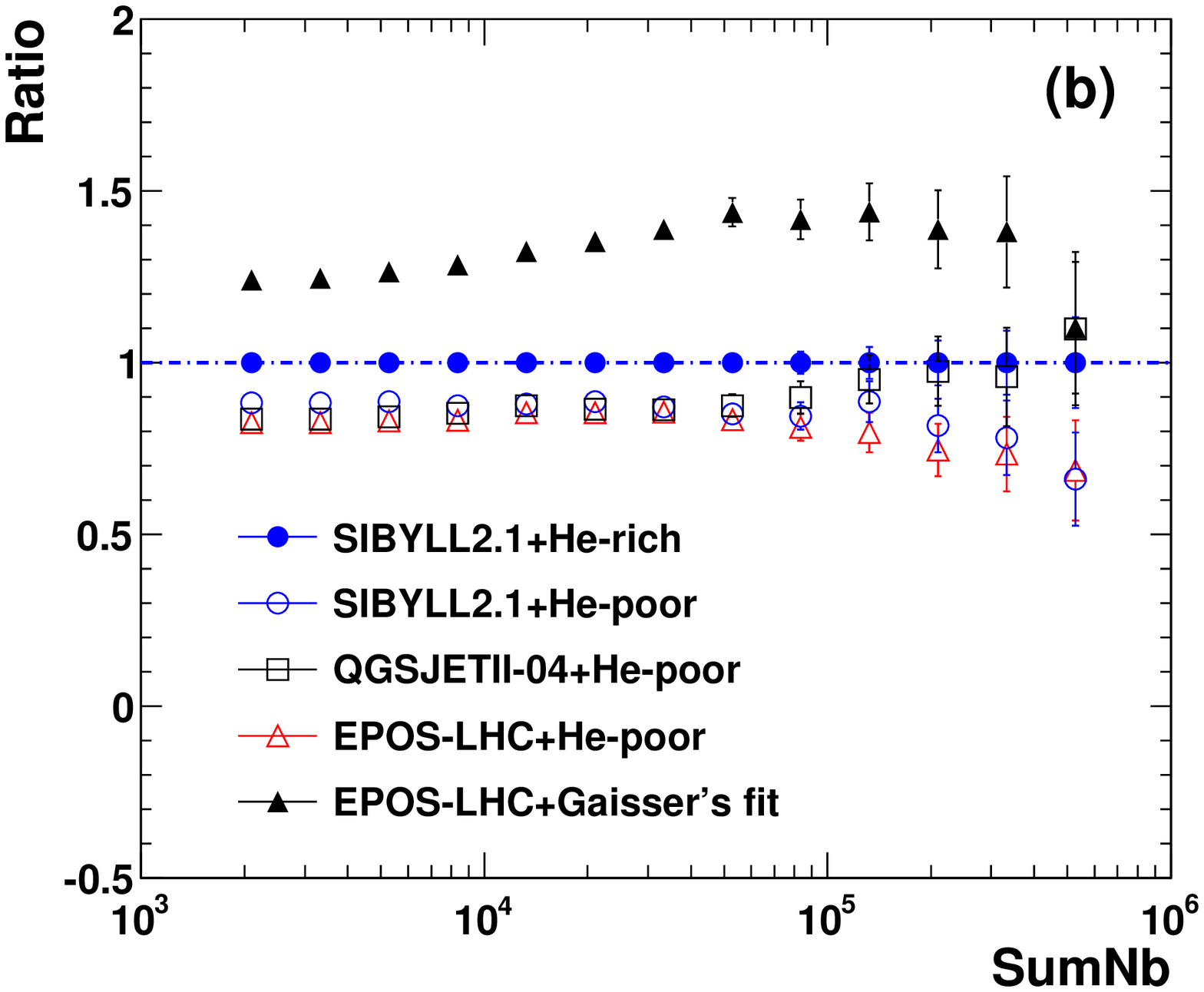}
  \includegraphics[width=0.48\linewidth]{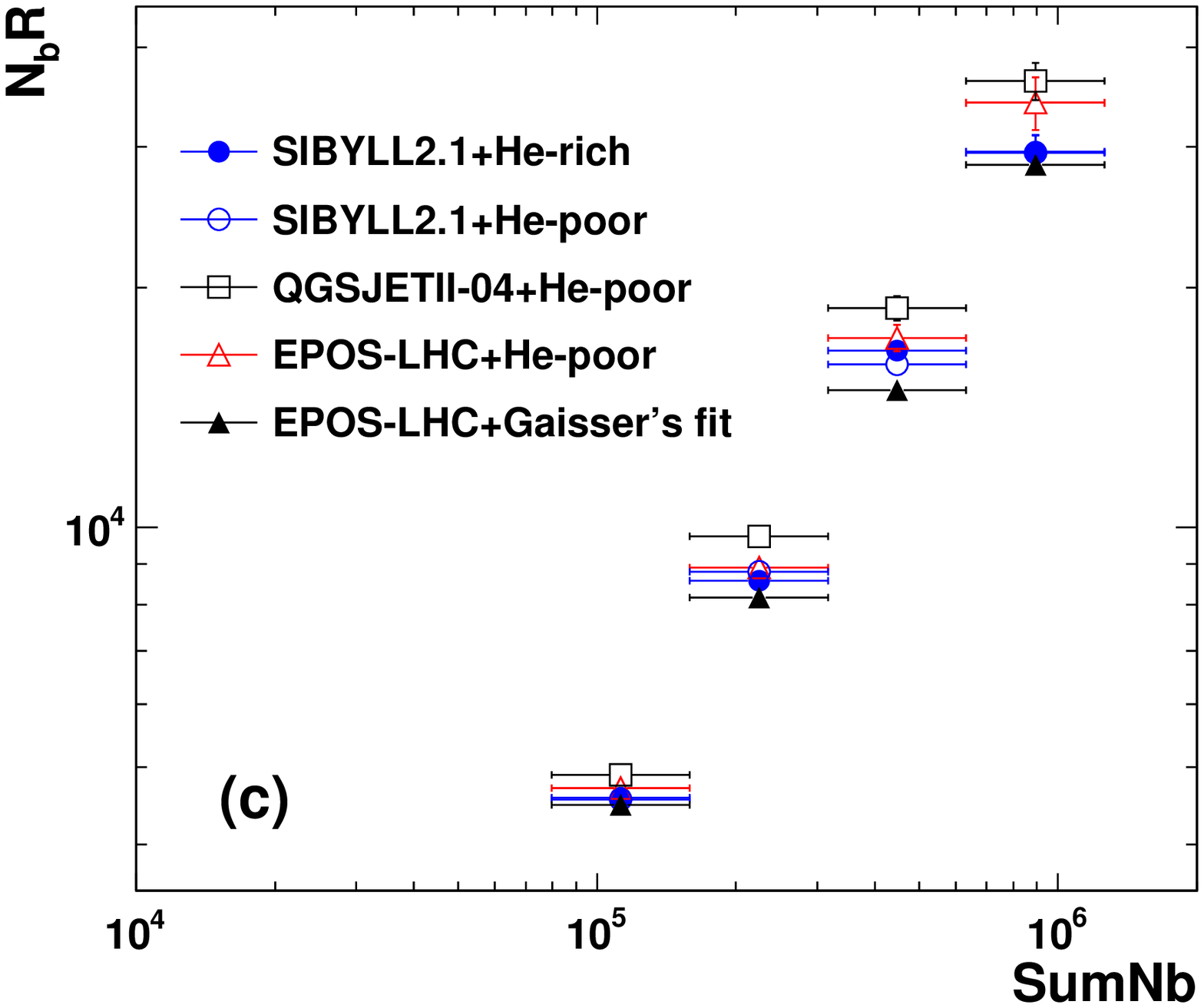}
  \caption{(a) Integral $\sum N_b$ (SumNb) spectra obtained from five MC models,   (b) The intensity ratios of $\sum N_b$ to that obtained by the SIBYLL2.1+He-rich model, (c) The mean energy-flow lateral spreads $\langle N_bR \rangle$ in the respective energy interval for
five MC models.}
  \label{figure-05}
 \end{figure}

It is well known that the absolute intensity of the total burst sizes depends sensitively on the increase of cross sections, inelasticity, and also on the primary cosmic-ray composition. Shown in Fig.~\ref{figure-05}-(a)  is  the  integral total burst-size spectrum ($\sum N_b$ (SumNb)) obtained by the respective  MC model for comparison. The $\sum N_b$ spectra obtained by five MC models
are compared each other by taking the flux ratio to that by the SIBYLL2.1+He-rich model in Fig.~\ref{figure-05}-(b).
It is seen that the EPOS-LHC+Gaisser-fit model gives the highest flux in  all $\sum N_b$ region. According to our  MC simulation, the observed AS cores in the size region of $\sum N_b$ = 2$\times 10^3$ - 4$\times 10^5$ are  produced mostly by the light component (P+He) with its primary energies of several times 10$^{14}$ - 10$^{15}$ eV.
The fraction of light component in the primary of this energy region is about 66\% for Gaisser-fit model while about 55\% for the He-rich model and 42\% for the He-poor model as seen in Table~\ref{table-1}. It should be, however, noted that about 70\% of the observed high-energy core events are originated by the light component, that is, the contribution from other nuclei is fairly small.

Here, we discuss the uncertainties due to different interaction models in reference to the spectrum of high-energy core events. Before entering in discussion, we should first remind that the production rate of high-energy core events is most sensitive to the energy per nucleon of primary particles. Thus, the mean energy per nucleon of helium nuclei is 1/4 of protons when compared at the same primary energy and also the interaction mean free path of helium nuclei in the air is about a half of that of protons. For the He-poor model, it is seen that the flux of protons is slightly  higher than that of helium nuclei or almost same in the energy region over about 300 TeV and also their power indices are almost same as learned from  Fig.~\ref{figure-03} and Table~\ref{table-1}. If we combine this with above discussion,
it  may be allowed to ignore the contribution  from helium nuclei to the core events observed in this primary model, that is, it is regarded as those produced by protons. Under this assumption, it may be noticed from Fig.~\ref{figure-05}-(b) that the flux values by SIBYLL2.1, QGSJETII-04 and EPOS-LHC using the same primary model of He-poor match well with an error of smaller than 10\%, which may be attributed to uncertainties due to the interaction models used. A deviation of QGSJETII-04 in the core-size region above about 2$\times 10^5$ may be due to a low statistics of the events, that is, within a statistical fluctuation (at most 2$\sigma$ level).

On the other hand, when we arrange the light component flux of protons and helium nuclei in descending order in the $10^2$ -$10^4$ TeV region, it becomes as Gaisser-fit $>$ He-rich $>$ He-poor. It is then confirmed that the flux of core events is in the same order as the primary light-component flux and also the shape of each primary spectrum is well reflected in the corresponding core-event spectrum, as seen in Fig.~\ref{figure-05}-(b). A typical example is seen in the case of Gaisser-fit primary model. This means that high energy core-event observation with YAC is very sensitive to the primary light-component spectrum.

A correlation between $\langle N_bR \rangle$ and $\sum N_b$ is shown in Fig.~\ref{figure-05}-(c).
This figure tells us that the Gaisser-fit primary model gives smaller lateral spread than others, while the QGSJETII-04+He-poor model gives larger spread than others.
As protons with the long interaction mean free path can penetrate deep in the atmosphere and produce AS cores near the observation level, resulting in giving smaller lateral spread. The Gaisser-fit
primary is light-component dominant as seen in Fig.~\ref{figure-03}, so that the core spread by this model should be smaller than others.
When the primary model is fixed as He-poor, the mean spread of QGSJETII-04 is slightly larger than that of SIBYLL2.1. This may slightly depend on the interaction model since the energy spectrum of secondary particles in the very forward
region (Feynman $x\sim$ 0.1 - 0.3) produced at collision in the SIBYLL2.1 model is harder than that of QGSJETII-04, that  is,  the former contains slightly larger number of very high-energy secondaries than the latter.  The difference of the intensity and lateral spread  between  both models could  be attributed  to the number of very high energy particles as those penetrate deep in the atmosphere.
It should be noted that the lateral spread of AS-cores is mostly caused by Coulomb scattering of shower electrons and positrons in the atmosphere, not by transverse momentum of secondaries produced at collisions except within the depth of 1-2 radiation lengths from  the interaction point in the atmosphere.
In connection with this, the energy loss of AS cores generated by QGSJETII-04 may be faster than those by SIBYLL2.1, resulting in giving lower AS-core intensity.
Hence, a  precise AS experiment like the (YAC + Tibet-III AS) array will be able to  examine the interaction models to some extent.

\subsection{Sensitivity of YAC-I to observe  the light-component spectrum around the knee}

In this analysis, we use the ANN technique to separate the light-component from others. This method is shown to be quite effective for such purpose as confirmed by our previous works~\cite{Tibet-EC-2000b,Tibet-EC-2006}.
In this ANN analysis, we use the following seven quantities :
 (1) $N_{hit}$, (2) $\sum$$N_b$, (3) $N_b^{top}$, (4) $\langle R \rangle$, (5) $\langle N_bR \rangle$, (6) $N_e$ and (7) $\theta$ (zenith angle).
These  are input to the ANN with 35 hidden nodes and 1 output unit. To train the ANN in separating light-component (P+He) from other nuclei, the input patterns for light-component and others are set to 0 and 1, respectively. We then define a critical value of $T_c$ to calculate the corresponding purity and selection efficiency of the selected (P+He)-like events.

 \begin{figure}[!ht]
  \centering
  \includegraphics[width=0.56\linewidth]{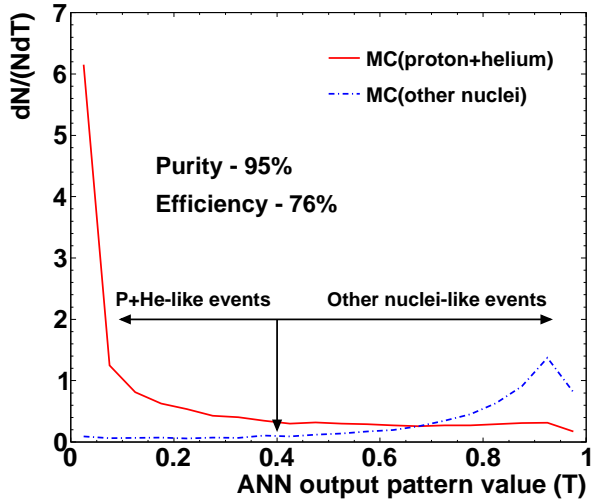}
  \caption{ ANN output pattern value (T) distribution of training (P+He) events based on the EPOS-LHC+He-poor model. The average selection purity and efficiency over whole energy range of (P+He)-like events are 95\%, 76\% at ${T_c}$ = 0.4.}
  \label{figure-06}
 \end{figure}

\begin{table}[!ht]
\caption{The ratios of (P+He)/All at three phases of analysis, before and after ANN training based on the MC models. In this table the second column represents the ratio of  primary (P+He) flux to the all-particle flux in the energy region above 40 TeV. The third column represents the ratio  of true (P+He) events contained in the observed high-energy core events selected by the condition mentioned in the text. The fourth column represents the ratio of true (P+He) events contained in the ANN trained core events with $T \leq T_c = 0.4$.  The fifth column represents  the ratio of the number of ANN trained core events  with the output $T \leq 0.4$ to that of all ANN trained core events ($0\leq T \leq 1$).}
\label{table-3}
\begin{indented}
\item[]
\begin{tabular}{@{}lcccc}
\br
%-------------------------------------------------------------------------------------------------------------------------------
  \multirow{2}{*}{Models}   &Primaries    &Core events  &\multicolumn{2}{c}{After ANN training}           \\
                            &(P+He)/all(\%)  &(P+He)/all(\%)  &Purity (\%)  &Efficiency (\%)       \\
\mr
%-------------------------------------------------------------------------------------------------------------------------------
  SIBYLL2.1+He-rich           &56.2     &$71.1\pm0.2$                    &$92.0\pm0.2$        &$70.1\pm0.3$     \\
  SIBYLL2.1+He-poor           &46.8     &$69.7\pm0.2$                    &$94.0\pm0.2$        &$71.1\pm0.3$     \\
  QGSJETII-04+He-poor         &46.8     &$69.5\pm0.3$                    &$94.6\pm0.2$        &$77.1\pm0.4$     \\
  EPOS-LHC+He-poor            &46.8     &$69.3\pm0.2$                    &$94.6\pm0.2$        &$75.8\pm0.3$     \\
  EPOS-LHC+Gaisser-fit        &67.0     &$87.4\pm0.1$                    &$96.1\pm0.2$        &$67.8\pm0.3$     \\
\br
%-------------------------------------------------------------------------------------------------------------------------------
\end{tabular}
\end{indented}
\end{table}

Figure~\ref{figure-06} shows the ANN output distribution trained using the  EPOS-LHC+He-poor model. As seen in this figure, the events with ${T_c} \leq 0.4$  could  be  regarded as the (P+He)-like events, and the average selection purity and efficiency over whole energy range of (P+He)-like events are 95\% and 76\%, respectively.
Table~\ref{table-3} is a summary of the ratios of the (P+He)/All before and after ANN training analysis based on
the five MC models over the whole energy range. Thanks to the performance of the YAC-I array, the ratio of the (P+He)/All before ANN training (core events) has already reached $\sim$70\% by the core-event selection conditions ($\sim$87\% for Gaisser-fit model, as seen in Table~\ref{table-3}).
With the ANN training, the purity of selected (P+He)-like events is further increased up to $\sim$95\% as learned from Table~\ref{table-3}.
This high quality data set is used for reconstructing the light-component primary spectrum.
Table~\ref{table-3} teaches us that after ANN training the difference of the selection purity and efficiency is within $\sim$6\% among three hadronic interaction models, and $\sim$8\% among three primary composition models.
Overall uncertainties due to ANN training for the MC models are then estimated to be about 10\%.

\subsection{Expected light-component spectrum}

\begin{figure}[!ht]
  \centering
  \includegraphics[width=0.56\linewidth]{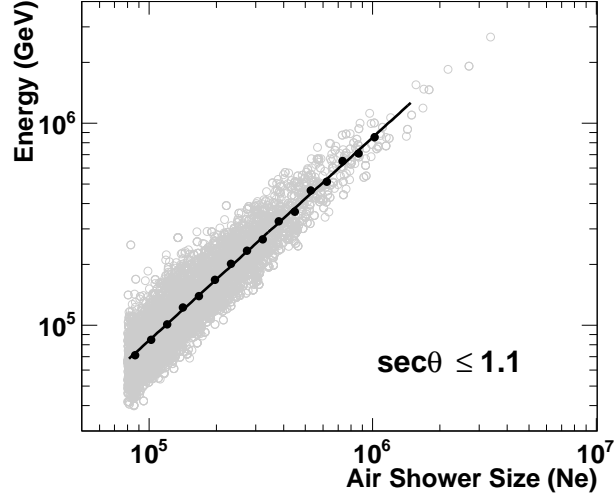}
  \caption{ Scatter plots of the primary energy ($E_0$) and the estimated shower size ($N_e$) of (Proton+Helium)-like events based on EPOS-LHC+He-poor model with $\sec\theta\leq1.1$. Solid line shows the fitting result of  $E_0$ = 0.794$\times$ ${N_e}^{1.005}$ GeV.}
  \label{figure-07}
 \end{figure}

Using the ANN trained by the EPOS-LHC+He-poor events, we select the (P+He) like events from all the observed events and also obtain the selection purity and efficiency. The primary energy $E_0$ of each selected events is then estimated
using the AS size $N_e$ obtained by the Tibet-III. The relation between air shower size $N_e$ and primary energy $E_0$ is expressed as

\[  E_0 = \alpha  \times N_e^\beta, \]
where the parameters of  $\alpha$ and $\beta$ are estimated from AS events generated by the MC  for the Tibet-III AS array, while $\alpha$ and $\beta$ depending on the zenith angle of air showers. Details of this procedure is described in the paper~\cite{Amenomori-2008}. Shown in Fig.~\ref{figure-07} is
the correlation between $E_0$ and $N_e$ of the (P+He)-like events selected by the ANN method. The  solid line denotes  the best fit curve for
this correlation, and the parameter values of $\alpha = 0.794$ and
$\beta = 1.005$ are then obtained for air showers with $\sec\theta \leq 1.1$.
The energy resolution is also
estimated as about 25\% at energies around 200 TeV.

We also checked a dependence of the correlation of $N_e$ and $E_0$ on the interaction and primary composition model, and obtained the similar relationship in the other models. We then found that there is less than 10\% difference for the determination of the primary energy by use of different interaction and primary composition models.

\begin{figure}[!ht]
  \centering
  \includegraphics[width=0.56\linewidth]{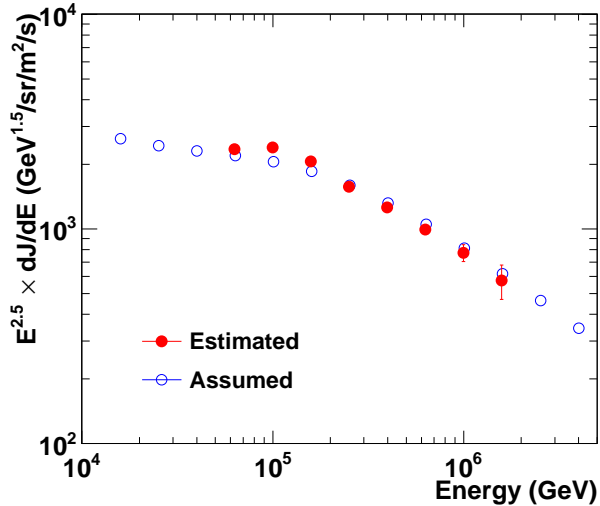}
  \caption{ The estimated energy spectrum of (Proton+Helium) compared with the assumed (input) one based on SIBYLL2.1+He-rich model. }
  \label{figure-08}
 \end{figure}

Figure~\ref{figure-08} shows the estimated primary  energy spectrum  of the
light-component (P+He) in comparison with the assumed primary spectrum.
It is seen that the estimated energy spectrum well reproduces the assumed one
within about 10\% errors. Almost same results are obtained for
other MC models.

In this work, we discuss the systematic errors coming from the models used in the MC simulation for deriving the primary (P+He) spectrum using the (YAC-I+Tibet-III) array. The systematic errors caused by each step of the analysis procedures are investigated including the dependence of the MC data on the interaction models, the primary composition models and the algorithms for the primary mass identification. We summarize these systematic errors as follows:
1) errors due to the observation efficiency ($S\Omega$ of (P+He)) depending on the interaction and primary composition models are found
to be smaller than 25\%;
2) errors due to the selection of (P+He)-like events with ANN training are about 10\%, in which the model dependence on the purity and efficiency is totally included;
3) errors due to the estimation of primary energy with use of the conversion from $N_e$ to $E_0$ are 10\%, in which the dependence on both interaction and primary composition models are taken into account;
4) errors due to the reconstruction procedures of the primary light-component spectrum from the observed core events are estimated to be smaller than 10\%.
The total systematic errors are then estimated to be about 30\% as the square root of quadratic sum of those four systematic errors, which may be somewhat overestimated because of a little correlation among four error estimation parts.

\section{Summary}

In this paper, we have carried out a full MC simulation to examine the capability of measuring the energy spectrum of
 primary light-component (P+He) of cosmic rays at the knee energies using
the (YAC-I+Tibet-III) array. The models used in this MC simulation are
SIBYLL2.1, EPOS-LHC (v3400) and QGSJETII-04 for the interaction and He-poor, He-rich and Gaisser-fit for the primary cosmic-ray composition. The Corsika code was used to generate AS events in the atmosphere and the  Geant4 code was used to
treat the shower particles entering in  the detectors. The air-shower core events
observed with the YAC-I array were analyzed to select those induced by the
light component using the ANN technique.
In this paper, we focused on the sensitivity of the YAC-I array to observe the light component of cosmic rays around the knee and discussed  the systematic errors coming from the the models used in the MC indispensable for obtaining the
result which should be  independent on the model as possible.
It is shown that the YAC+Tibet AS array is powerful to study the primary cosmic-ray chemical composition, in particular, to obtain the energy spectrum of light-component (P+He) of cosmic rays at the knee energies.
A full-scale YAC consisting of 400 core detectors covering more than 5000 m$^2$ could be operated
 together with the Tibet-III array in the very near future.

\ack
The authors would like to express their thanks to the members of the Tibet AS$\gamma$ collaboration for the fruitful discussion. This work is supported by the Grants from the National Natural Science Foundation of China (Y11122005B, Y31136005C and Y0293900TF) and the Chinese Academy of Sciences (H9291450S3) and the Key Laboratory of Particle Astrophysics, Institute of High Energy Physics, CAS. The Knowledge Innovation Fund (H95451D0U2 and H8515530U1) of IHEP, China also provide support to this study.


\section*{References}

\begin{thebibliography}{00}


\bibitem{Amenomori-2008} Amenomori M \etal 2008 {\it Astrophys. J. } {\bf 678} 1165

\bibitem{Horandel-2003} H\"{o}randel J R 2003 {\it Astropart. Phys. } {\bf 19} 193

\bibitem{Horandel-2004} H\"{o}randel J R 2004 {\it Astropart. Phys. } {\bf 21} 241

\bibitem{Shibata-2010} Shibata M \etal 2010 {\it Astrophys. J. } {\bf 716} 1076

\bibitem{Tibet-EC-2000a} Amenomori M \etal 2000 {\it Phys. Rev. D } {\bf 62} 072007

\bibitem{Tibet-EC-2000b} Amenomori M \etal 2000 {\it Phys. Rev. D } {\bf 62} 112002

\bibitem{Tibet-EC-2006} Amenomori M \etal 2006 {\it Phys. Lett. B } {\bf 632} 58

\bibitem{ATIC2-2005} Wefel J P \etal 2005 {\it Proc. of 29th Int. Cosmic Ray Conf. } (Pune, India, 3-10 Aug. 2005) vol~3 p~105

\bibitem{CREAM-nuclei} Ahn H S \etal 2009 {\it Astrophys. J. } {\bf 707} 593

\bibitem{CREAM-PHe} Yoon Y S \etal 2011 {\it Astrophys. J. } {\bf 728} 122

\bibitem{YAC-detector-jiang} Jiang L \etal 2009 {\it Proc. of 31st Int. Cosmic Ray Conf. } ({\L}\'{o}d\'{z}, Poland, 7-15 Jul. 2009)

\bibitem{Hardware-YAC-Liwj} Amenomori M \etal 2011 {\it Proc. of 32nd Int. Cosmic Ray Conf. } (Beijing, China, 11-18 Aug. 2011) vol~3 p~301

\bibitem{Hardware-YAC-chen} Amenomori M \etal 2011 {\it Proc. of 32nd Int. Cosmic Ray Conf. } (Beijing, China, 11-18 Aug. 2011) vol~3 p~290

\bibitem{Heck-Corsika} Heck D \etal 1998 {\it Forschungszentrum Karlsruhe Report FZKA 6019 }

\bibitem{siby} Engel R \etal 1999 {\it Proc. of 26th Int. Cosmic Ray Conf. } (Salt Lake City, USA, 1999) vol~1 p~415;
               Ahn Eun-Joo \etal 2009 {\it Phys. Rev. D} {\bf 80} 094003

\bibitem{eposlhc} Pierog T \etal 2013 arXiv:1306.0121[hep-ph]

\bibitem{qgs} Ostapchenko S 2011 {\it Phys. Rev. D } {\bf 83} 014018

\bibitem{Gaisser-model} Gaisser T K 2012 {\it Astropart. Phys. } {\bf 35} 801

\bibitem{JACEE-1998} Asakimori K \etal 1998 {\it Astrophys. J. } {\bf 502} 278

\bibitem{RUNJOB-2001} Apanasenko A V \etal 2001 {\it Astropart. Phys. } {\bf 16} 13

\bibitem{Geant4} Agostinelli S \etal 2003 {\it Nucl. Instrum. Meth. Phys. Res. A } {\bf 506} 250

\bibitem{ANN} Peterson C \etal 1994 {\it Comp. Phys. Comm. } {\bf 81} 185







\end{thebibliography}
\end{document}